\documentclass[twoside,11pt]{article}
\usepackage{amsmath,amsfonts,amssymb,verbatim,float,anysize,fancyvrb,graphicx,multicol,mathrsfs,mdframed}
\usepackage[utf8]{inputenc}
\usepackage[doublespacing]{setspace}
\usepackage[english]{babel}
\RequirePackage[colorlinks,citecolor=blue,urlcolor=blue]{hyperref}
\usepackage[numbers]{natbib}
\usepackage[usenames,dvipsnames, svgnames]{xcolor}
\usepackage{authblk}
\usepackage{multicol}
\usepackage{times}
\usepackage{subfig}
\usepackage{graphicx}
\usepackage{graphicx}
\graphicspath{{figures/}}
\usepackage{booktabs}
\usepackage[font=small,labelfont=bf]{caption}
\usepackage{amsfonts, amsmath, amsthm, amssymb}
\usepackage{wrapfig}
\usepackage[margin=2cm]{geometry}
\definecolor{mygreen}{rgb}{0,0.6,0.5}
\definecolor{myblue}{rgb}{0,0.45,0.7}
\definecolor{myred}{rgb}{0.8,0.4,0}
\definecolor{mygray}{rgb}{.6,.6,.6}

\parskip = \baselineskip
\newlength{\normalparindent}
\AtBeginDocument{\setlength{\normalparindent}{\parindent}}

\begin{document}
\setlength\parindent{24pt}

\title{Flexible Bayesian Inference on Partially Observed Epidemics}

\author{Maxwell H Wang \thanks{email: maxwang@hsph.harvard.edu}}
\author{Jukka-Pekka Onnela\thanks{email: onnela@hsph.harvard.edu}}
\affil{\small{Harvard TH Chan School of Public Health}}

\maketitle

\begin{abstract}
{Individual-based models of contagious processes are useful for predicting epidemic trajectories and informing intervention strategies. In such models, the incorporation of contact network information can capture the non-randomness and heterogeneity of realistic contact dynamics. In this paper, we consider Bayesian inference on the spreading parameters of an SIR contagion on a known, static network, where information regarding individual disease status is known only from a series of tests (positive or negative disease status). When the contagion model is complex or information such as infection and removal times is missing, the posterior distribution can be difficult to sample from. Previous work has considered the use of Approximate Bayesian Computation (ABC), which allows for simulation-based Bayesian inference on complex models. However, ABC methods usually require the user to select reasonable summary statistics. Here, we consider an inference scheme based on the Mixture Density Network compressed ABC (MDN-ABC), which minimizes the expected posterior entropy in order to learn informative summary statistics. This allows us to conduct Bayesian inference on the parameters of a partially observed contagious process while also circumventing the need for manual summary statistic selection. This methodology can be extended to incorporate additional simulation complexities, including behavioral change after positive tests or false test results.}
{Network; Contagion; Bayesian Statistics}\\
\end{abstract}

\section{Introduction}

In the study of infectious diseases, mathematical models are useful for predicting trajectories of outbreaks or evaluating the efficacy of intervention strategies. The spread of disease is dependent on non-random, heterogeneous patterns of human contact. To capture the realistic dynamics of interpersonal interactions, it is sometimes necessary to move beyond traditional fully-mixed models to individual-based or agent-based models and incorporate contact network information. Contact networks provide a natural representation of human mixing patterns, where a population of individuals are represented as nodes and potential transmission contacts are represented as edges. By leveraging contact network structure, it is possible to infer transmission paths of past epidemics \citep{groendyke2012}, identify individuals vital to the spread of contagion \citep{pastor2002, lu2016}, or propose strategies that modify contact network topology in order to control disease \citep{gates2015}. Contact networks have also been extended to non-disease contagions, such as the spread of behavior and misinformation \citep{shao2018}. Although network data has traditionally been difficult to obtain, emerging technologies, such as Bluetooth proximity sensing, make it increasingly feasible to obtain this type of information at scale.

When considering inference on parameters governing spreading processes on networks, it is appealing to consider statistical inference from a Bayesian perspective. Under this framework, parameters of interest are treated as random variables with an initial prior distribution and the target of inference is the distribution of these parameters conditioned on the observed data. The Bayesian perspective allows for the incorporation of prior information on uncertain parameters from domain experts, while also providing transparent and interpretable results in the form of posterior distributions. However, in the study of infectious disease, data is commonly missing. Oftentimes, complete history describing the evolution of an epidemic is unavailable. Infection and recovery times may not be precisely recorded and need to be inferred solely from observations of disease status at specific time points. For some individuals, the disease status may be missing altogether, perhaps due to lack of testing or absence of symptoms. Thus, there exists a need for network methods that incorporate the uncertainty in real-world data. 

Previous work on Bayesian inference for network epidemics has often focused on methods that sample directly from an analytical posterior via Monte Carlo Markov Chain (MCMC) algorithms \citep{oneill2009, groendyke2011, bu2022, schweinberger2022}. In these methods, uncertainty in the data is accounted for via data augmentation, where missing information is treated as a set of latent variables that are then jointly inferred upon along with parameters of interest. However, when large amounts of data are missing, data-augmentation methods can become computationally infeasible due to the high dimensionality of the latent space. Furthermore, in order to maintain analytical feasibility, it may be necessary to limit the expressiveness and flexibility of the model. As models become more complex, MCMC methods often require the design of specialized, problem-specific algorithms.

Approximate Bayesian Computation (ABC) describes a set of simulation-guided methods that allow for direct sampling from an approximate posterior distribution without the need to specify a likelihood. Such methods can be used to conduct inference on the dynamics of contagions in both homogeneously mixed populations \citep{mckinley2009, numminen2013, kypraios2017} and on contact networks \citep{walker2010, dutta2018, almutiry2020}. Due to its flexibility, ABC has also found use in other applications in genetics, ecology, and physics \citep{tishkoff2001, toni2009, akeret2015}. Under an ABC framework, parameter values are sampled from the prior distribution and applied to a model. Parameter values that lead to simulated data deemed suitably similar to the observed data are accepted as samples from the approximate posterior.

As ABC methods can suffer from the curse of dimensionality, it is often necessary to summarize the observed data with low-dimensional summary statistics \citep{blum2010}. Outside of exponential likelihood families, Bayes sufficient statistics are usually unavailable. Thus, dimension-reducing summary statistics must be provided by the user, often based on scientific or intuitive understanding of what metrics of epidemic evolution may be relevant to the problem. There are numerous techniques for identifying summary statistics that are both low-dimensional and highly informative. One group of such techniques aims to choose a best subset from an existing set of summary statistics \citep{joyce2008, nunes2010, raynal2019}, though such methods typically require a proposed set of user-defined summary statistics to select from. Other methods transform a given set of statistics to construct lower-dimensional statistics \citep{blum_and_francois2010, fearnhead2012, prangle2014, jiang2017}. In \cite{hoffmann2022}, it is shown that many such methods are special cases or large-sample limits of, or equivalent to, an information-theoretic approach that minimizes expected posterior entropy.

In this paper, we investigate the use of the Mixture Density Network-compressed ABC (MDN-ABC) for disease parameter inference on an incompletely observed epidemic. The MDN-ABC uses a conditional mixture density network to learn low-dimensional but informative summary statistics by minimizing the expected posterior entropy \citep{hoffmann2022}. Unlike the MDN itself, however, the MDN-ABC makes no parametric assumptions about the true posterior. Thus, this framework enjoys the modeling flexibility and asymptotic guarantees of ABC methods while also eliminating the need to define or select summary statistics.

In Section 2 of this paper, we will define the notation used to describe a contagious process on a network. We will also discuss the basics of ABC algorithms and the architecture used for the MDN-ABC. In Section 3, we will investigate the performance of MDN-ABC on a simple, fully-observed SI epidemic and demonstrate its fidelity to existing closed-form solutions. In Section 4, we will demonstrate the use of MDN-ABC on a partially observed SIR process where disease state of individual nodes is known only from asynchronous tests. We will consider a range of synthetic networks as well as an empirical social network from Karnataka, India \citep{banerjee2013}.

\section{Methods}

\subsection{Approximate Bayesian Computation}
In Bayesian inference, parameters are treated as random variables. Inferences are based on the posterior distributions: the probability distribution of parameters conditioned on the observed data. This posterior distribution takes the form of 
\begin{equation}
\pi(\theta|Y) \propto L(Y|\theta)\pi(\theta).
\end{equation}
Here, $L(Y|\theta)$ is the likelihood of the data and $\pi(\theta)$ is the prior distribution of the parameter, which can be used to include information regarding previous understanding of the parameter. When both the likelihood and prior are tractable, there exist various methods for sampling from the posterior distribution. However, in complex models, these expressions may be difficult to specify in closed form. This includes many stochastic simulation-based problems, as even simple mechanistic models can have complex likelihoods.

Approximate Bayesian Computation (ABC) is a likelihood-free method first named by \cite{beaumont2002}, though similar methods had previously been applied to problems in population genetics \citep{fu1997, tavare1997}. Crucially, ABC does not require a tractable likelihood. It requires only that given a proposed parameter value, a dataset can be forward-simulated from the model. The simulated data is then compared to the observed data. Parameter values that produce data deemed similar enough to the observed data are accepted as posterior samples. The typical rejection ABC follows a simple rejection sampling scheme:
\begin{enumerate}
  \item[1.] Sample a candidate parameter value $\theta'$ from prior $\pi(\theta)$.
  \item[2.] Forward simulate the model using $\theta'$ to obtain a simulated datset $Y'$.
  \item[3.] Given distance function $d$, compute distance between simulated and observed datasets: $d(Y_{obs}, Y')$.
  \item[4.] Consider acceptance threshold $\epsilon$. If $d(Y_{obs}, Y') \leq \epsilon$, accept $\theta'$ as a sample from the posterior. 
  \item[5.] Return to Step 1 until a predetermined number of samples are obtained.
\end{enumerate}

ABC is approximate in two aspects. First, as $\epsilon \rightarrow 0$, the ABC posterior converges to the target posterior. However, even when data is discrete, setting $\epsilon = 0$ usually leads to prohibitively large numbers of simulations to draw a desired number of posterior samples. In the continuous case, $P(Y_{obs} = Y) = 0$. Thus, it is necessary to set $\epsilon > 0$. In such situations, the ABC algorithm is not drawing samples from the proper posterior, but an approximation. 

In addition, the observed raw data $Y_{obs}$ is often high-dimensional. As the kernel smoothing implicit in ABC is subject to performance degradation in high-dimensional settings, it is often necessary to instead calculate summary statistics $S(Y_{obs})$ and $S(Y')$. Steps 3 and 4 are then replaced with:
\begin{enumerate}
  \item[3a.] Given distance function $d$, compute distance between simulated and observed datasets: $d(S(Y_{obs}), S(Y'))$.
  \item[4a.] Consider acceptance threshold $\epsilon$. If $d(S(Y_{obs}), S(Y')) < \epsilon$, accept $\theta'$ as a sample from the approximate posterior.
\end{enumerate}

When both of these approximations are applied, the distribution being approximated by ABC is
\begin{equation}
p(\theta|s_{obs}) = \int p(\theta,s|s_{obs})\, \mbox{d}s \propto \int K_\epsilon(||s-s_{obs}||)p(s|\theta)p(\theta) \, \mbox{d}s.
\end{equation}
Here, $s_{obs} = S(Y_{obs})$, $||u||$ is the Euclidean norm of $u$, and $K_\epsilon(||u||) = K(\frac{||u||}{\epsilon})/\epsilon$ is a smoothing parameter \citep{blum2010}. The rejection ABC draws samples from the joint conditional distribution $p(\theta,s|s_{obs})$, which serves as a good approximation to the true posterior when $\epsilon$ is small and the summary statistic $s_{obs}$ is highly informative.

There exists a rich body of literature on other ABC sampling algorithms that can improve computational performance. Examples include the Monte Carlo Markov Chain ABC \citep{marjoram2003} and the Sequential Monte Carlo ABC \citep{sisson2007}. Such methods typically allow for more efficient sampling; instead of sampling repeatedly from a potentially uninformative posterior, the proposal region from which the parameter values are drawn is gradually narrowed down. In this paper, we consider only the basic rejection ABC algorithm for the sake of simplicity. However, our methodology affects only the definition of summary statistics $S(Y)$, so extensions to other ABC algorithms are straightforward.

\subsection{MDN-Compressed ABC}

\begin{center}
\begin{figure}[H]
\includegraphics[page=1, width=15cm]{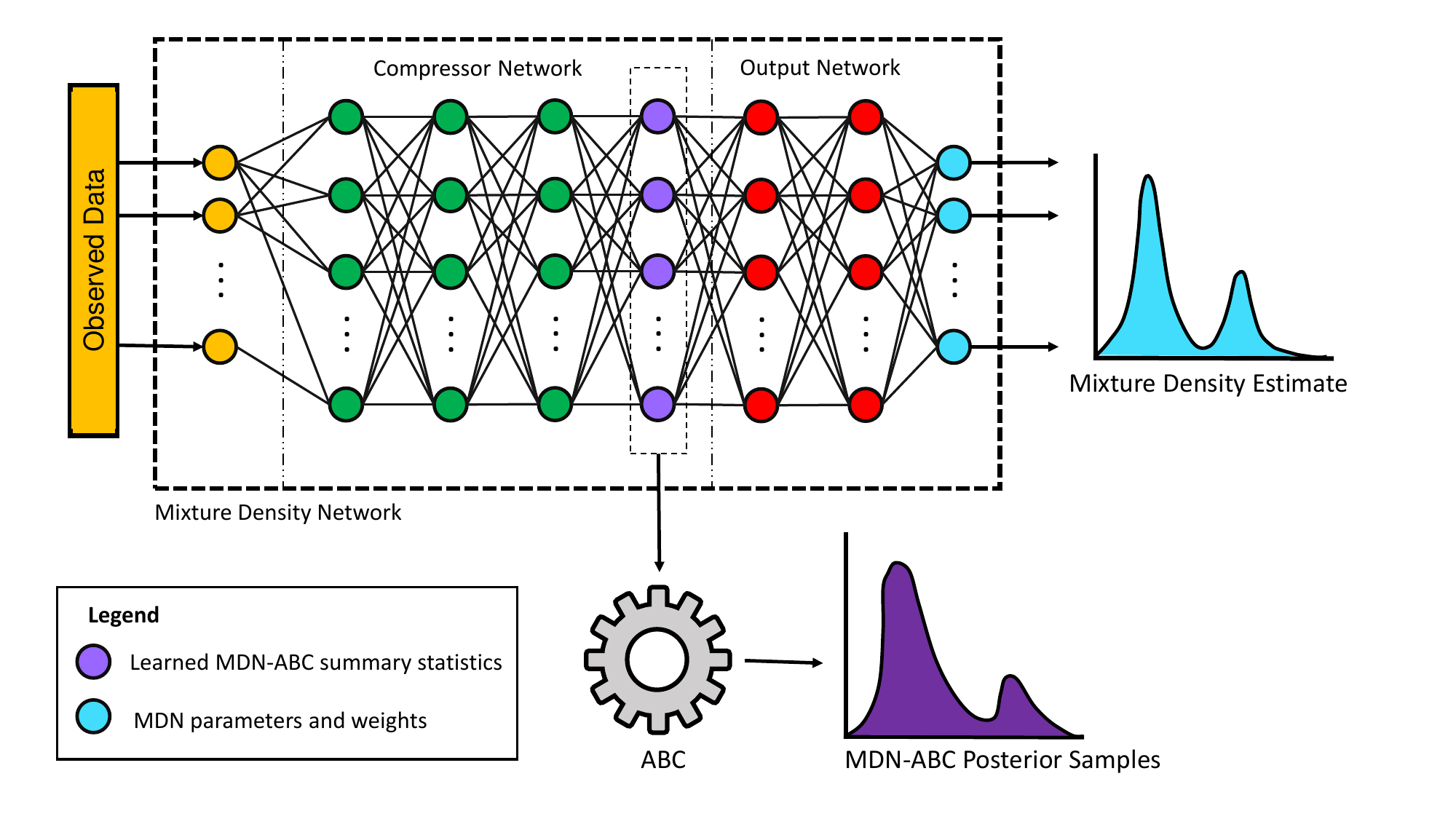}
\caption{The MDN-ABC utilizes two simple neural networks in series. The combined neural network inside the dashed rectangle is trained to generate a conditional mixture density that minimizes the expected posterior entropy (right). However, the estimated mixture density relies on parametric assumptions about the posterior distribution. Instead, we extract a single layer of the full neural network (the output layer of the compressor network) and treat it as our summary statistic for ABC. This schematic is a simplified visualization, and the depth and width of the neural networks utilized may not correspond to those pictured.}
\label{mdn_abc_schematic}
\end{figure}
\end{center}

In order to avoid the curse of dimensionality, it is usually necessary to compress the output data of our model via some statistic $s(Y)$. However, in general, Bayes sufficient statistics, such that $p(\theta|s(Y_{obs})) = p(\theta|Y_{obs})$, are not available. Thus, it is necessary instead to opt for optimal summary statistics that minimize some loss functional. \cite{hoffmann2022} proposed minimization of the expected posterior entropy (EPE)
\begin{equation}
\mathcal{H} = -\int p(s,\theta) \mbox{log}p(\theta|s) \, \mbox{d}s \, \mbox{d}\theta.
\end{equation}
Typically, the loss function employed for neural network training is the Monte Carlo estimate of the expected posterior entropy:
\begin{equation}
\hat{\mathcal{H}} = -m^{-1}\sum_{i=1}^m \mbox{log}f(\theta_i, s(Y_i)),
\end{equation}
where $f(\theta,t)$ is a conditional density estimator that approximates the posterior, $\theta_i$ and $Y_i$ are joint samples from $p(\theta,Y)$, and $m$ is the number of samples in the minibatch. The minimization of the EPE is equivalent to other information theoretic approaches, including minimizing expected Kullback-Leibler divergence, maximizing mutual information, and maximizing expected surprise. 

Conceptually related work has proposed the use of mixture density networks (MDNs) \citep{bishop1994} to learn a conditional posterior density estimation that minimizes the EPE \citep{papamakarios2016}. The MDN itself is a combination of a neural network and a mixture density model. Using the Monte Carlo estimate of the EPE as the loss function, the neural network is trained to learn the weights and the parameters for the mixture distribution that approximates the posterior. For example, if the mixture distribution takes the form of $K$ Gaussian components, the neural network of the MDN would learn the $K$ mixture weights, means, and variances associated with the mixture distribution. Compared to ABC, MDN methods have the advantage of directly learning an approximate distribution to the posterior, instead of simply sampling from an $\epsilon$-ball centered on the observed data \citep{papamakarios2016}. However, such methods rely on parametric assumptions about the posterior distribution and do not enjoy the same asymptotic guarantees as ABC.

\cite{hoffmann2022} introduce a method dubbed MDN-compressed ABC, or MDN-ABC. This framework combines MDN and ABC methods using two simple neural networks combined in series. The overall network is trained to minimize EPE with a mixture density network. Instead of using the output of the MDN, the ABC summary statistics are extracted as the output of the first neural network (known as the compressor network). The second neural network (known as the output network) is utilized only for the learning of the conditional density estimator. A schematic of this method is shown in Figure \ref{mdn_abc_schematic}.

MDN-ABC learns highly informative summary statistics, and compares well with the MDN posterior. Furthermore, in some scenarios, the posterior may be multimodal or otherwise complex. While the conditional density estimate can theoretically approximate the posterior arbitrarily closely given infinite computational budget, the parameteric assumptions may be too restrictive to capture the true posterior in practical situations. Other methods that estimate the posterior mean of the parameters as the summary \citep{fearnhead2012} also have difficulty with multimodal posteriors. However, the MDN-ABC can still perform well, due to the lack of parametric assumptions. In this paper, we will focus on the MDN-ABC; however, MDN-ABC yields a conditional density estimator as a byproduct, so our framework can be easily extended to a pure MDN approach.

\subsection{Epidemics on Networks}

We consider a network $\mathcal{G}$. $\mathcal{G}$ consists of a set of nodes $\mathcal{N}=\{1,...,n\}$, representing individuals, which are connected by edges $\mathcal{E}\subseteq \mathcal{N}\times\mathcal{N}$. In studies of epidemics, such edges represent potential paths of transmission. In this paper, we focus on static, undirected, unweighted networks. However, the ABC framework allows for flexible implementation of more complex network dynamics as well. As long as the model can be easily forward-simulated, such complexity does not affect the asymptotic properties of ABC.

We consider two types of contagion, expressed through compartmental models \citep{kermack1927}. In an SI process, we assume that nodes can take on one of two disease states: susceptible (S) and infectious (I). Nodes begin in the susceptible state, and progress to the infectious state with a per-contact rate of $\beta$ in continuous time. This parameter is directly translatable to the definition of transmissibility in \citep{newman2002}. Once infected, nodes are capable of transmitting the infection to susceptible contacts, and they remain the infectious state for the remainder of the epidemic. In an SIR process, nodes can take on an additional state: recovered (R). Under this model, infected nodes progress to the recovered state with  rate $\gamma$. Once recovered, nodes can no longer infect susceptible nodes or be infected by infected nodes.

We consider continuous SIR epidemics similar to those defined in \citep{fennell2016}. At time $t$ each node $i \in \mathcal{N}$ has a disease state $X_t^i$. For SI models, $X_t^i \in \{S,I\}$, and for SIR models, $X_t^i \in \{S,I,R\}$. The instantaneous transition rates are defined as follows:
\begin{eqnarray}
\gamma & = & \lim_{\Delta t\to 0} \frac{P(X_{t+\Delta t}^i = R | X_t^i = I)}{\Delta t} \\
\beta & = & \lim_{\Delta t\to 0} \frac{P(X_{t+\Delta t}^i = I \mbox{ via } j | X_t^i = S, X_t^j = I)}{\Delta t}.
\end{eqnarray}
Here, $\{X_{t+\Delta t}^i = I \mbox{ via } j\}$ is the event that node $i$ is infected by infected neighbor $j$. Note that for SI epidemics, $\gamma = 0$. In this paper, we will primarily focus on the posterior distributions of $\beta$ and $\gamma$ as the targets of inference. 

It has been shown that coarse discretization of continuous-time processes can lead to misleading conclusions \citep{fennell2016}. Thus, we implement a continuous-time Gillespie simulation of an SIR epidemic.

\begin{enumerate}
  \item[1.] Begin with $B< |\mathcal{N}|$ nodes in the infected state. All other nodes are considered susceptible. Initiate at $t = 0$.
  \item[2.] While $t < t_{max}$:
      \begin{itemize}
      \item[2.1] For each node $i$, calculate a transition coefficient $\alpha_i$ based on node $i$'s current status.
      	\begin{itemize}
        	\item[a)] Recovered nodes have transition rate 0.
            \item[b)] Infected nodes have transition rate $\gamma$.
        	\item[c)] Susceptible nodes have transition rate $m_i\beta$, where $m_i$ is the number of node $i$'s neighbors that are currently in the infected state.
    	\end{itemize}
       \item[2.2] Increment $t$ by random variable $\tau$, where $\tau \sim \mbox{Exponential}(\frac{1}{\sum_i\alpha_i})$.
       \item[2.3] Select a single node $j$ to transition. The probability of selection for each node $i$ is $\frac{\alpha_i}{\sum_i \alpha_i}$.
       \item[2.4] If node $j$ is susceptible, it becomes infected. If node $j$ is infected, it becomes recovered.
       \item[2.5] If there are no remaining infected nodes, end simulation. Otherwise, return to Step 2.
      \end{itemize}
\end{enumerate}

For this simple infection process, if infection and recovery times are fully observed for all nodes and gamma-distributed priors are assumed for $\beta$ and $\gamma$, closed-form solutions are available for the maximum likelihood estimates and the posterior distributions of both parameters: \citep{bu2022}. However, stochastic network epidemics, even when mechanistically simple, can have complex likelihoods. The complexity is further compounded when data for infection and recovery times are missing. Thus, this specific problem is a good candidate for ABC methods.

\section{Simple Example: Fully Observed SI Epidemic}
In this section, we will consider a completely observed SI epidemic and compare the results of the MDN-compressed ABC to closed-form solutions. Here, $\beta$ is the only parameter we wish to infer on, as $\gamma = 0$. Since all infection times ${e_j}$ are observed, the scenario is simple enough for closed-form solutions to be available.
Derived in \citep{bu2022}, the maximum likelihood estimate for $\beta$ is:

\begin{equation}
\hat{\beta}_{MLE} = \frac{n_I - 1}{\sum_{j=1}^n e_{SI}(t_j)(t_j - t_{j-1})}
\end{equation}

Here, $n_I$ is the total number of infected individuals at the end of the epidemic. $e_{SI}(t)$ is the number of edges existing between susceptible and infected individuals at time $t$, and $t_1 < t_2 < ... < t_n$ are ordered infection times for all nodes.

Next, given a Gamma-distributed prior for $\beta$ such that $\beta \sim \mbox{Gamma}(a,b)$, the posterior distribution of $\beta$ given ordered event times $t_j$ is:
\begin{equation}
\beta|{t_j} \sim \mbox{Gamma}(a + (n_I-1), b + \frac{(n_I-1)}{\hat{\beta}_{MLE}})
\end{equation}
In order to diagnose sensitivity of our method to network topology, we varied two aspects of the underlying contagion network: degree distribution (Poisson or log-normal) and mean degree (2, 4, or 8). This led to six  scenarios, for all of which MDNs were trained using an identical neural network architecture, optimizer, and training regime. Poisson-distributed networks are generated as Erdős–Rényi random graphs \citep{erdos1960} and log-normal distributed networks were generated with a Chung-Lu model \citep{chung2002} with expected degrees drawn from a log-normal distribution with $\sigma^2$ fixed at $0.5$. We also ensured that each graph consisted of a single connected connected component. For each component outside of the largest connected component (LCC), we selected a random unconnected dyad between the LCC and the smaller component and added one additional edge. All networks consisted of 100 nodes.

We choose the prior for $\beta$ to be $\mbox{Gamma}(2, 4)$, a distribution that is much more disperse (variance of 0.125) than the true posterior and has a mean (0.5) different than the true value of $\beta$. We set the true value of $\beta$ to be 0.15 and chose a single origin node to be infected at time $t = 0$. The simulation continues until all nodes reach the infected state. 

For each network scenario, a total of $5\times10^6$ realizations of the continuous-time SI simulation were generated for the training set, and $2.5\times10^6$ realizations were generated for the validation set. For efficient computation, such simulations can be generated completely in parallel. Each simulation is run by drawing a proposal $\beta$ from the Gamma prior and forward-simulating the epidemic to obtain the infection time of each node. Every simulation uses the same underlying network and begins at the same origin node selected in the original epidemic. Methods for inferences on an unknown source are discussed in \citep{dutta2018}.

For the MDN-ABC, the raw data is the infection time $t_i$ for each node. The MDN is trained by minimizing the expected posterior entropy with a similar method to \citep{hoffmann2022}. We utilize two gamma components for our conditional posterior density, and a 15-dimensional feature space. Our compressor is a simple feed-forward neural network with 4 hidden layers. The total number of nodes per layer is: $[100,80,60,30,20,15]$. All layers are fully connected with hyperbolic tangent activation functions. The output network consists of three fully-connected neural networks with 2 hidden layers (nodes per layer: $[15,10,10,2]$) that learns the mixture weights, shapes, and rates of the two gamma-distributed components for the MDN. The entire neural network was trained with stochastic gradient descent with a minibatch size of 500. We used an Adam optimizer with the learning rate initialized at $5\times10^{-5}$. After each epoch (full pass through the training data), validation loss was evaluated by calculating the Monte Carlo estimate of EPE on the validation set. If 10 epochs elapsed without an improvement in validation loss, training was terminated.

\begin{figure}[H]
\begin{center}
\includegraphics[page=1, width=12cm]{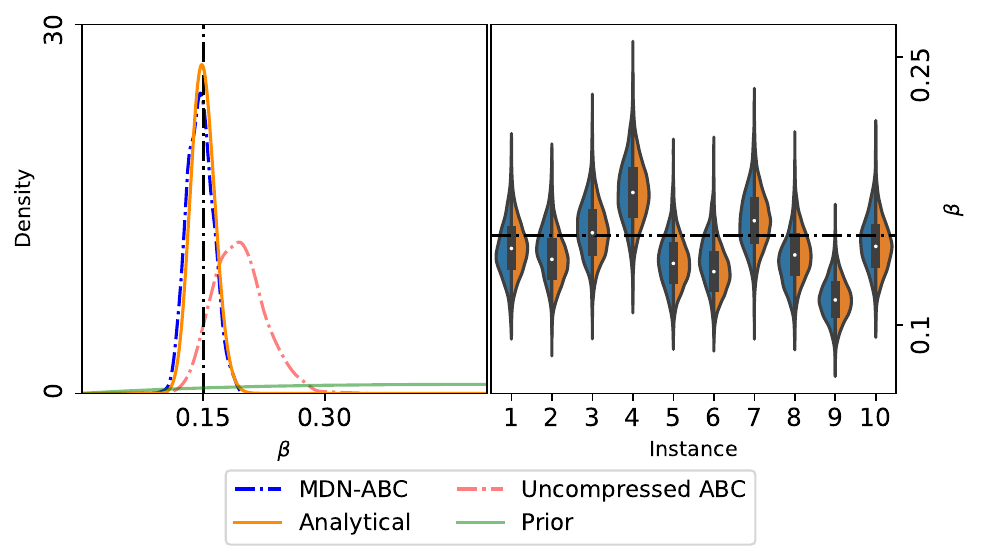}
\caption{The left figure shows the Gaussian kernel density estimate (used for visualization purposes only) of the posterior draws from the MDN-ABC (blue) and rejection ABC from the uncompressed raw data itself (red). The right figure shows violin plots comparing the MDN-ABC posteriors to the gold standard, across 10 difference instances of the epidemic. The underlying network is log-normal with a mean degree of 8. }
\label{twin_plot}
\end{center}
\end{figure}

For our ABC, we reused the $5\times10^6$ training realizations as our pool of simulations. Using the summary statistics learned by our MDN-ABC architecture, we calculated the Euclidean distance between each the 10-dimensional summary statistic vector of each simulation output and the 10-dimensional summary statistic calculated from the original ``observed" epidemic. We then selected the best $0.02\%$ of the training set, resulting in approximately 1000 posterior samples. In addition, we also drew posterior samples using the raw data itself, $(t_1, \ldots, t_{100})$, as the statistic of interest, terming this the ``uncompressed ABC". For a single network scenario (log-normal degree distribution with mean degree 8), the Gaussian kernel estimates of the posteriors drawn from the compressed MDN-ABC and the uncompressed ABC are shown in Figure \ref{twin_plot} (left), where they are compared with the closed-form posterior from \cite{bu2022}. 

In addition, the epidemic trajectory of the original ``true" epidemic is stochastic. In order to examine the variance in posteriors introduced by this stochasticity, we regenerate 10 instances of ``original" epidemics and redraw 1000 MDN-ABC posterior samples without retraining the MDN. In Figure \ref{twin_plot} (right), we compare these 10 MDN-ABC posteriors to the gold standard closed-form solutions. 

\begin{figure}[H]
\centering
\includegraphics[page=1, width=4in]{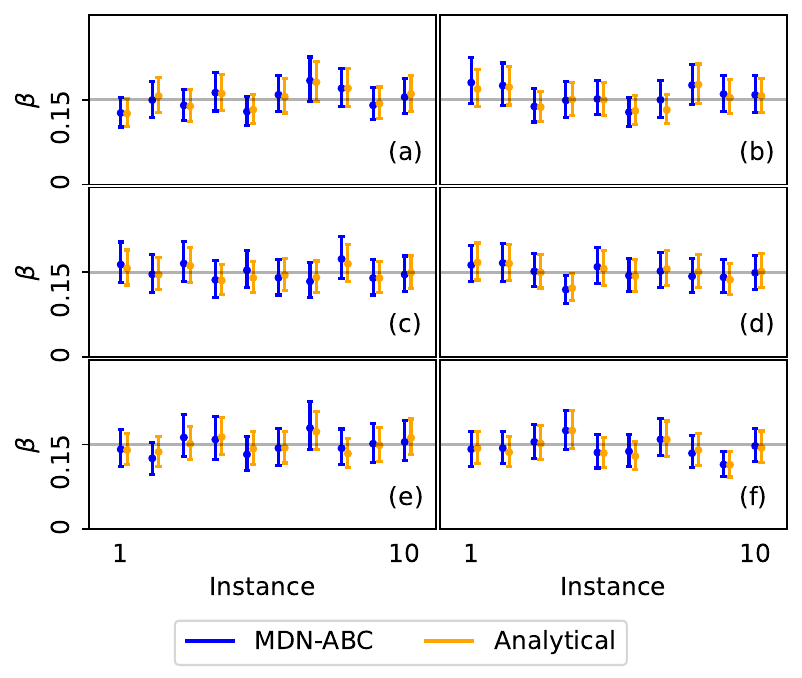}
\caption{Comparison of 95\% credible intervals for posterior samples from the MDN-ABC to samples from the analytical solution, over six network scenarios: a) Poisson with mean degree 2, b) log-normal with mean degree 4, c) Poisson with mean degree 4, d) log-normal with mean degree 4, e) Poisson with mean degree 8, and f) log-normal with mean degree 8, each with 10 instances of an original epidemic.} \label{interval_comparison}
\end{figure}

Similarly, for every network scenario, we generated 10 instances of original epidemics and used MDN-ABC to obtain 1000 posterior samples. In Figure \ref{interval_comparison}, we show $95\%$ credible intervals obtained from the MDN-ABC samples, compared with 5000 samples drawn the closed-form solution for the posterior distribution.

In this example, we can observe that the MDN-ABC can generate posterior samples that match closely with the analytical solution. However, in more complex situations, such as scenarios where the infection and recovery times are missing, closed-form solutions are not known. In addition, due to the missing event times, many summary statistics such as those employed in \cite{almutiry2020} and \cite{dutta2018}, including the time of the peak of the epidemic curve, the mean of infection times, and the length of epidemic, cannot be calculated. In the next section, we consider the use of MDN-ABC in this case of the partially observed epidemic.

\section{SIR Epidemic with Missing Outcome Data}
Previous literature regarding inferences on network epidemics have typically assumed that event times are precisely known \citep{dutta2018, almutiry2020}, or are aggregated over coarse time intervals \citep{walker2010, bu2022}. In previous ABC applications to network epidemics, the chosen summary statistics, such as the time of the peak of the epidemic \citep{almutiry2020} or the proportion of nodes infected at each time step \citep{dutta2018}, are dependent on exact knowledge of event times.

However, in most real world epidemics, the exact times of infection and recovery are often unobserved. Instead, information regarding disease status is known from observations of individual disease status. These observations may not be synchronous, as the disease status of individuals may be observed at different times. Some individuals may not be observed at all. As the test status of individuals is not necessarily observed at the time of their transition between states, coarse aggregations of outcomes (e.g. number of infections per day) may also be misleading. Similar work by \citep{kypraios2017} has also considered ABC for applications in which case-detection times of an epidemic are observed; however, this work makes the assumption that case-detection corresponds to removal time, and all such removal times are detected. In our example, we do not assume that individuals who have received a positive test proceed to follow any quarantine procedure, but our model can easily be extended to accommodate test-dependent contact avoidance.

Similarly to the fully-observed SI epidemic, for the SIR epidemic with missing outcome data, we chose the prior for $\beta$ and $\gamma$ both to be $\mbox{Gamma}(2, 4)$. We set the true value of $\beta$ to be 0.15 as before and the true value of $\gamma$ to be 0.1. We initialize $5\%$ of the population to be infected at time $t = 0$. The simulation continues until time $t=50$. Unlike the fully-observed scenario, exact infection and recovery times are not available. Instead, each node is randomly assigned a time $0 < t < 7$ to begin testing and is then tested every 7 time steps. If each time step is considered to be a day, this would correspond to a testing cadence of one week. A test returns positive if the node is infected, and returns negative if the node is susceptible or recovered. In this paper, we consider mandatory tests on individuals, such that an individual's probability of being tested or adhering to a testing schedule is independent of their disease status. Tests are also assumed to have perfect sensitivity and specificity, though false positives and negatives can easily be incorporated into the model. All contact networks consist of 100 nodes in one connected component. Over the simulation spanning $t=50$ time steps, the 100 nodes yield a total of 720 test results. 

A total of $5\times10^6$ realizations of the continuous-time SI simulation were generated for the training set, and $2.5\times10^6$ realizations were generated for the validation set. The raw data consists of the sequence of test results (positive or negative) obtained for each node. Again, the MDN learns two gamma components for our conditional posterior density. Our compressor consists of 6 hidden layers ($[720, 300, 200, 100, 60, 40, 30, 15]$), and the output network has 2 hidden layers ([$15,15,15,6$]). We extract a total of 15 summary features. The MDN is trained with the same training regime as the fully-observed SI model. MDN-ABC posterior samples are again drawn by picking the best $0.02\%$ of the training data, minimizing the Euclidean distance in the 15-dimensional summary statistic vector.

\begin{figure}%
    \centering
    \subfloat{{\includegraphics[width=4.5in]{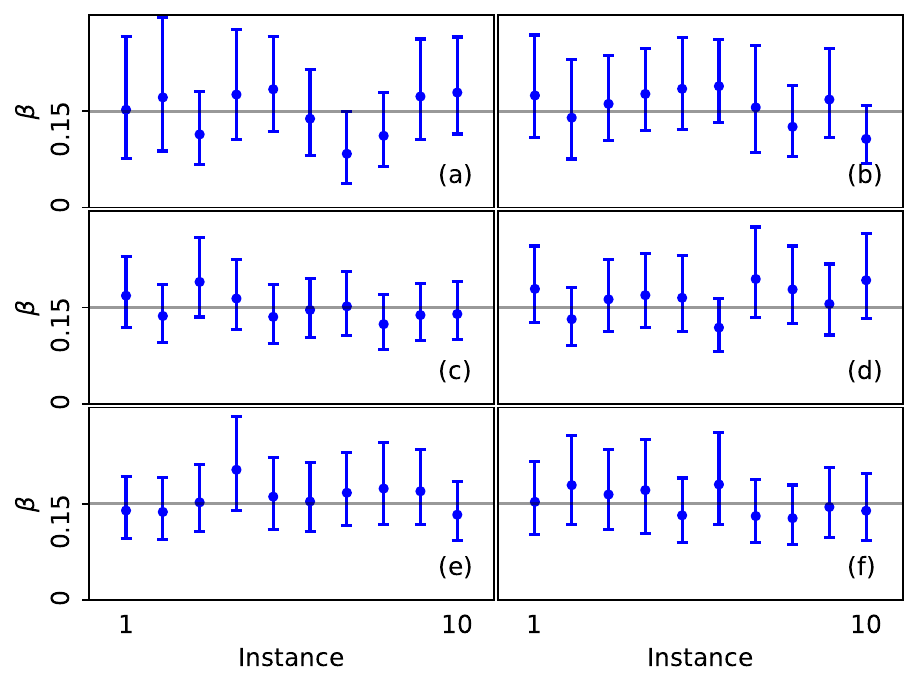} }}%
    \qquad
    \subfloat{{\includegraphics[width=4.5in]{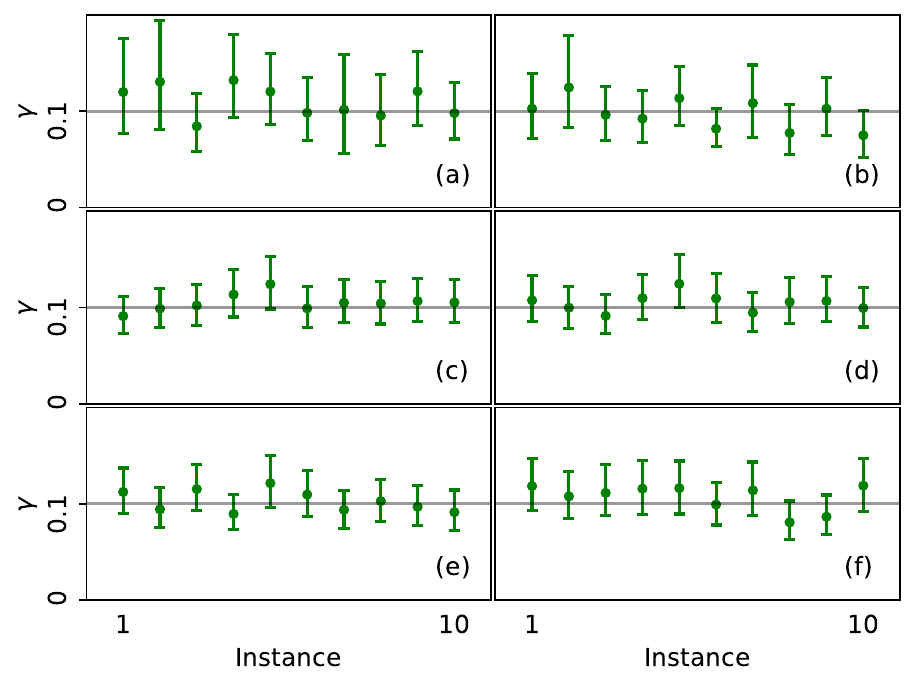} }}%
    \caption{Visualization of 95\% credible intervals for posterior samples from the MDN-ABC for a partially observed SIR epidemic for 10 realizations of the original epidemic, over six network scenarios: a) Poisson with mean degree 2, b) log-normal with mean degree 4, c) Poisson with mean degree 4, d) log-normal with mean degree 4, e) Poisson with mean degree 8, and f) log-normal with mean degree 8.}%
    \label{intervals}%
\end{figure}

\begin{figure}%
    \centering
    \subfloat{{\includegraphics[width=2.2in]{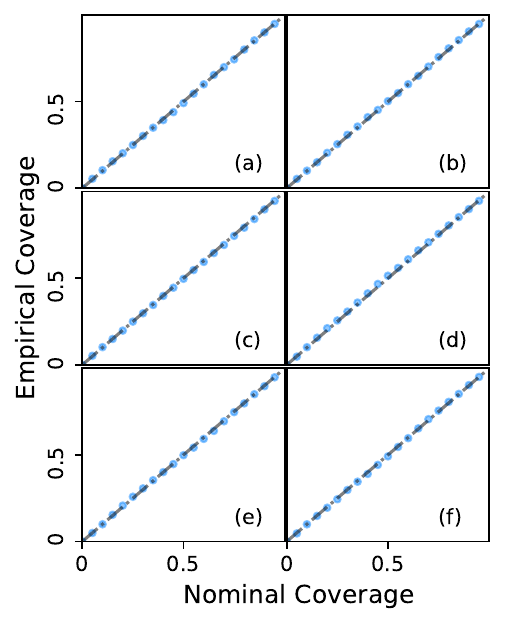} }}%
    \qquad
    \subfloat{{\includegraphics[width=2.2in]{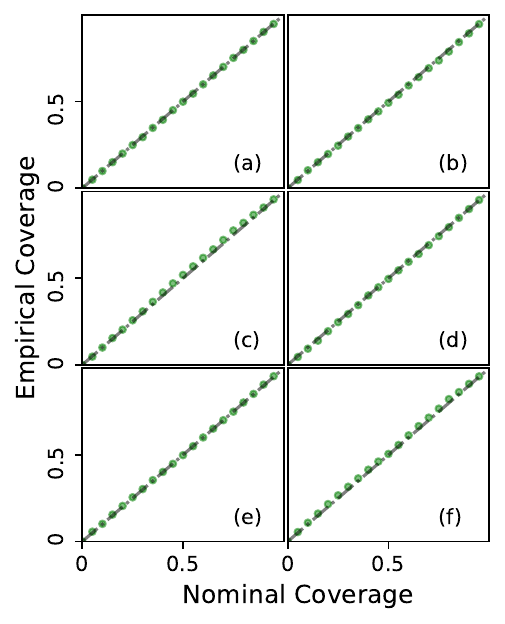} }}%
    \caption{Coverage properties for MDN-ABC posteriors for a partially observed SIR epidemic, over six network scenarios: a) Poisson with mean degree 2, b) log-normal with mean degree 4, c) Poisson with mean degree 4, d) log-normal with mean degree 4, e) Poisson with mean degree 8, and f) log-normal with mean degree 8.}%
    \label{coverages}%
\end{figure}

Again, to examine sensitivity of our method to varying degree distributions, we plot 95\% credible intervals obtained across six different scenarios for the underlying contact network (Poisson and log-normal distributed degree distributions, with mean degrees of 2, 4, and 8) in Figure \ref{intervals}. For mass action models, the intensity of an epidemic is often summarized with the basic reproduction rate $R_0$, which describes the expected number of secondary infections arising from a single infected individual introduced to a completely naive homogeneous population. Such calculations are more complicated in network models with heterogeneous contact structures. To develop intuition about the expected severity of epidemics, we consider the expected number of cases arising from a single infection introduced into a susceptible network. As derived in \citep{volz2009}, if each edge has a uniform probability of transmission $\tau$ and the initial infected node is randomly selected from the population, the expected number of infections arising from that node is $\tau C_1$ and the expected number of infections arising from infected nodes other than the first is $\tau \frac{C_2}{C_1}$. Here, if $k = \{0,...,n-1\}$, and $p_k$ is the degree distribution of the network,
\begin{equation}
    C_1 = \sum k p_k
\end{equation}
\begin{equation}
    C_2 = \sum (k-1)kp_k
\end{equation}
We calculate these values for the empirical degree distributions of the synthetic networks, each consisting of 100 nodes, generated for our six network scenarios. For the Poisson-distributed network with approximate mean degree 2, $C_1 = 2.30$ and $C_2 = 4.18$. For the Poisson-distributed network with approximate mean degree 4, $C_1 = 4.24$ and $C_2 = 16.92$. For the Poisson-distributed network with approximate mean degree 8, $C_1 = 7.80$ and $C_2 = 59.96$. For the log normal-distributed network with approximate mean degree 2, $C_1 = 2.30$ and $C_2 = 5.52$. For the log normal-distributed network with approximate mean degree 4, $C_1 = 3.98$ and $C_2 = 20.50$. For the log normal-distributed network with approximate mean degree 8, $C_1 = 8.38$ and $C_2 = 106.72$. As one example, setting $\tau =1$, a randomly selected initially infected node in the Poisson-distributed network with mean degree 4 would lead to $C_1 = 4.24$ expected secondary infections and subsequent infected nodes would each lead to $\frac{C_2}{C_1} = \frac{16.92}{4.24}=3.99$ expected secondary infections.

In order to validate our posterior samples, we also consider the coverage properties of the credible intervals derived from MDN-ABC samples \citep{cook2006, prangle2014_diagnose, talts2018}. For each network scenario, we drew tuples $(\beta_{(1)},\gamma_{(1)}),\allowbreak ..., (\beta_{(n)},\gamma_{(n)})$ from the prior distributions of $\beta$ and $\gamma$. Using these values of $\beta$ and $\gamma$, we simulated $n=5000$ epidemics and used MDN-ABC to obtain 5000 sets of posterior samples. Following from \citep{prangle2014_diagnose}, we define an $\alpha\%$ credible interval as the interval $I$ such that $\mbox{Pr}(\theta\in I|Y_{obs})=\alpha/100$. If the MDN-ABC posterior samples are a good approximation to the true posterior, the $\alpha\%$ credible intervals estimated from the MDN-ABC posterior samples should contain the ``true" values of $\beta$ and $\gamma$ in $\alpha\%$ of the simulations. Thus, for each of the 5000 simulated epidemics, we used the credible intervals estimated from each set of MDN-ABC posterior samples to calculate the empirical coverage for for $\alpha\%$ intervals ranging from $\alpha=0$ to 100. The empirical coverage plots are shown in Figure \ref{coverages}. Significant departures from linearity would indicate a poor fit, while linearity indicates that that MDN-ABC provides a good approximation to the true posterior.

\subsection{Empirical Network: Karnataka}
To demonstrate the use of MDN-ABC on a real-world network, we simulated an SIR process on a social network sampled from a village in Karnataka, India \citep{banerjee2013}. The network consists of 354 nodes representing individual people of whom 346 are part of the largest connected component (LCC). Edges represent relationships between individuals, and were obtained from surveys in which individuals named others with whom they interacted with. Individuals were specifically questioned regarding a number of social interaction types, including transaction of money, exchange of advice, and home visits. The network is assumed to be undirected, as relationships are assumed to be reciprocal. The mean degree is approximately 8.7, and the maximum degree was 35.
The simulated epidemic is identical to the one described in the previous section. The true value of $\beta$ was set to 0.15 and the true value of $\gamma$ to 0.1. To seed the epidemic, 5\% of the population was initialized in the infected state. All individuals were assigned a day of the week to begin testing, and infection statuses are known only from weekly tests that return a positive or negative status. The prior for each parameter was a Gamma(2,4) distribution. For the MDN, we employed a similar neural network architecture as the partially observed SIR case, though we now extract 20 summary statistics as opposed to 15. 

\begin{figure}
    \centering
    \includegraphics[width = 6in]{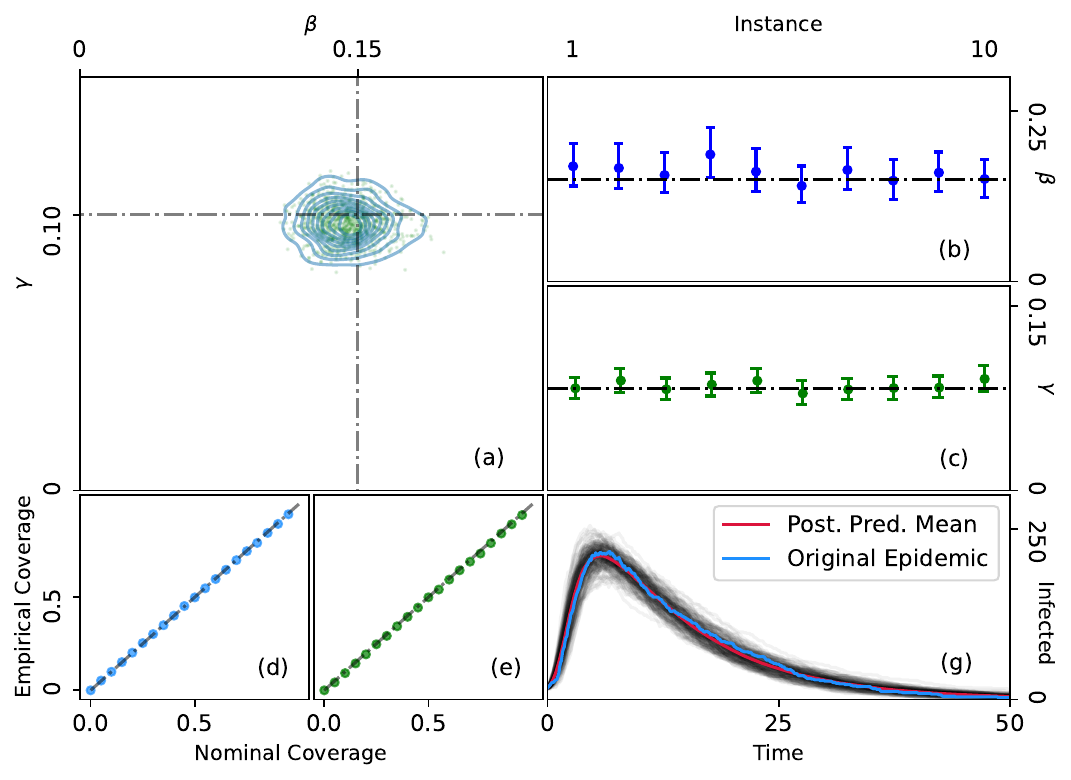}
    \caption{MDN-ABC results for simulated epidemic on Karnataka village network: a) MDN-ABC posterior samples, b) 95\% credible intervals for $\beta$ across 10 instances of original epidemic, c) 95\% credible intervals for $\gamma$ across 10 instances of original epidemic, d) empirical coverage for $\beta$, e) empirical coverage for $\gamma$, and f) 150 epidemic trajectories drawn from the posterior predictive.}
    \label{karnataka}
\end{figure}

In Figure \ref{karnataka}, we display the results for MDN-ABC on a the simulated epidemic on the Karnataka network. We also once again extract the 95\% credible intervals across 10 different realizations of the original epidemic and examine the coverage properties by drawing independent realizations of $(\beta, \gamma)$ from the prior and evaluating the empirical coverage probabilities of the MDN-ABC posteriors. Lastly, we re-simulate epidemics using values of $\beta$ and $\gamma$ drawn from the MDN-ABC posterior, sampling from the posterior predictive of epidemic trajectories. 

\subsection{Interpretability of Features}
While MDN-ABC allows users to skip the subjective summary statistic selection phase of ABC, this task is delegated to a black-box neural network. The individual summary statistics are a layer of neurons within the neural network trained on simulated datasets. Unfortunately, the meaning of each individual statistic produced by the MDN often lacks the interpretability of more intuitive measures of epidemic progression, such as statistics describing the epidemic curves and subgraphs induced by infected individuals. 

In the field of neural networks, especially in image recognition, rich literature exists on interpretability for neural network behavior. Such methods include saliency maps \citep{adebayo2018} and activation minimization \citep{erhan2009}. These techniques can be applied to qualitatively interpret MDN-ABC features. However, optimization-based visualization may struggle in this specific example. Given a particular epidemic model and a fixed network, not all epidemic trajectories are possible. For example, in our model, nodes receive the contagion from infected neighbors (there are no outside sources of infection); thus, no node can become infected before its neighbors have been infected, or after all of its neighbors have recovered. 

For one example of feature visualization, we created a synthetic network similar to a Cayley tree. The origin node of the epidemic at the center of the tree is parent to 6 subtrees, where each subtree is a complete binary tree with 4 layers. We again generate a partially observed SIR epidemic, this time using $\beta = 0.35$ and $\gamma = 0.07$ for a more aggressive epidemic. The testing cadence was once again $t=7$ time steps, and the simulation proceeded for a total of 50 time steps. We employed a similar architecture from the partially-observed SIR epidemic simulation example, and trained an MDN with 2 gamma components and 15 features. We then extracted the training samples that generated the maximum activation in each of the 15 neurons that serve as the MDN-ABC features (it may also be useful to extract the minimum. Below, we display the infected nodes and times of infection for the six training epidemics that contribute to the greatest activation of the ninth neuron out of 15 (recovery times are not pictured). By using such visualizations, it becomes possible to visualize what epidemic trajectories are generally represented by each feature. 

Notably, the data used to train the neural network did not represent the entire epidemic trajectory; the only information available for training was the series of tests results, positive and negative, observed during the epidemic. However, the epidemics represented by these training samples share topological features that are relatively interpretable by humans. For example, the epidemics that led to the maximum activation of the ninth MDN-ABC feature exhibited relatively rapid spread, indicating that the underlying $\beta$ is large. In addition, these epidemics were largely contained to a single subtree in the network, suggesting that the infectious period is also short (i.e. the origin node recovered before passing the contagion onto the remaining subtrees).

\begin{figure}[H]
\begin{center}
\includegraphics[page=1, width=5in]{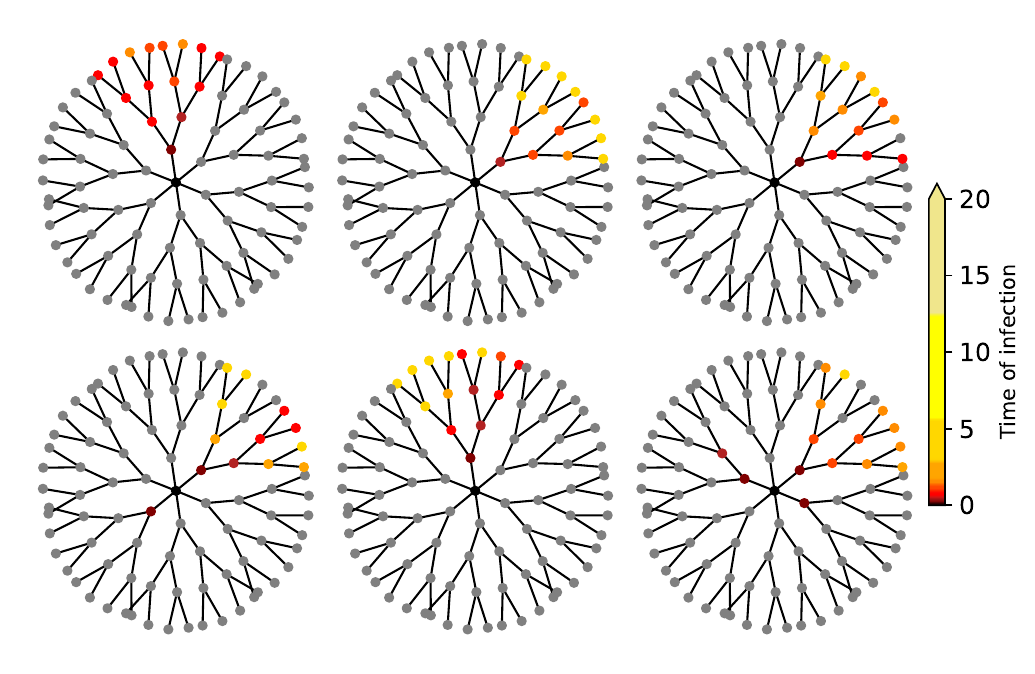}
\caption{Six training samples that yield the maximum activation in feature neuron 9 of 15. Shades of nodes denote time of infection. Grey nodes remained uninfected during the epidemic.} \label{block}
\end{center}
\end{figure}

\section{Discussion}
This paper presents MDN-ABC as a method for conducting Bayesian inference for network epidemics in situations where information for outcomes comes only from waves of reports of disease status. As summary statistics are no longer defined by the user but learned by an MDN via EPE minimization, MDN-ABC offers increased flexibility in handling diverse data types and allows for increased model expressiveness in simulation-based inference. 

Our paper invites several directions for extension. Our work primarily focuses on relatively simple compartmental disease models with the SI and SIR contagion types. Potential extensions to more complex compartmental models such as SEIR may better describe specific diseases of interest, such as COVID-19 and influenza. In addition, most networks are not closed systems, and infection can enter the network from an outside community. For such situations, the incorporation of an additional ``spark term", an instantaneous contact-independent rate of infection, may be more realistic. Furthermore, the current model assumes that an individual's test/observation schedule is independent of their disease status. This may be a reasonable assumption for cases where tests are mandated, which was a common intervention against COVID-19 \citep{lanier2021, karthikeyan2021, earnest2022}. This assumption may also hold for diseases observed in wildlife \citep{powell2020} or livestock, as well as for environmental testing such as testing on wastewater \citep{karthikeyan2021}. However, in many realistic situations, individuals may be more likely to seek out testing if they experience symptoms or if their close contacts are known to have been infected. Similarly, in realistic situations, knowledge of a positive test may cause individuals to cease contact momentarily with others, effectively removing them from the study population \citep{o1999, hambridge2021}. These additional layers of complexity can be added to the simulated model to better reflect a particular epidemic of interest and could be explored in future work. 

The underlying contact networks in our simulations are assumed to be unweighted, undirected, static, and perfectly observed. The likelihood-free nature of ABC allows for the inclusion of more complex contagion-network interactions, so implementation of weighted edges or dynamic networks is straightforward. Another avenue for further work is in the incorporation of uncertainty in the observed network itself. Typically, networks must be imputed from observed data, such as contact diaries and surveys \citep{de1978, mastrandrea2015} or close-proximity events detected by wearable RFID devices \citep{cattuto2010, mastrandrea2015} or Bluetooth \citep{stopczynski2014, huang2016}. Accounting for the uncertainty associated with the underlying contact network is an important next step for Bayesian inferences on network epidemics. Methods similar to \cite{young2020} may be combined with MDN-ABC to sample from the joint posterior of the epidemic parameters of interest and parameters that model the missingness of network data. This approach can also incorporate statistical network models such as Exponential Random Graph Models \citep{robins2007} and Congruence Class Models \citep{goyal2014}.

One drawback of the MDN-ABC is its significant computational cost compared to other ABC approaches. It requires the training of a mixture density network, so sufficiently large datasets for training and validation must be simulated. However, these simulations represent independent realizations of epidemics, so this problem is embarassingly parallel. In addition, once training of the neural network is complete, more advanced ABC sampling algorithms can be applied for increased sampling efficiency while utilizing the summary statistics defined by the MDN.

In addition, for the purposes of this paper, we utilize a simple feedforward neural network architecture for our MDN. However, recent advances in neural network models may be employed for more efficient learning of the mixture density parameters. For example, developments in graph neural networks (GNNs) allows users to leverage the underlying relationships within the data \citep{scarselli2008}. Typically, GNNs employ a message passing scheme that allows nodes to gather information about neighboring nodes via a chosen aggregation operator. For network epidemics, test results or event times may be represented as node-level attributes. Extensions to GNNs include heterogeneous graph neural networks \citep{zhang2019}, which allow for the inclusion of different types of nodes with different attributes. In this paper, GNNs are not strictly necessary to obtain reasonable results. Thus, we choose to use a relatively generic feedforward neural network, which was intuitively simple and easily applicable to both epidemic settings that we discussed (fully observed SI process and partially observed SIR process). This simpler neural network architecture may also be more easily generalizable to epidemic models that may not necessarily have network-structured data, such as compartmental models or agent-based models. However, in real-world applications of MDN-ABC, it may be beneficial to explore more specialized neural network architectures for the specific type of epidemic model and available data.

While network data has traditionally been difficult to obtain, emerging technologies, such as Bluetooth proximity sensing, make it increasingly feasible to obtain this type of information at scale. The availability of contact network information can be used to study infectious diseases with realistic and high-resolution models. However, there remains a need to incorporate real-world uncertainties in inferences on contagion parameters. By minimizing EPE, it is possible to employ an MDN to learn informative summary statistics for ABC settings, while allowing for uncertainty associated with partial observations. MDN-ABC can thus be used as a flexible method to develop a more nuanced understanding of infectious disease spread and contribute to individual-level risk exposure assessments and targeted interventions.

\section{Acknowledgements}
This work was supported by the National Institutes of Health [T32AI007358, R01 AI138901]. We would also like to thank Victor De Gruttola, Ravi Goyal, Till Hoffmann, and the members of the Onnela Lab for their constructive feedback.

\bibliography{citations}

\end{document}